\documentstyle[eqsecnum,aps]{revtex}

\begin{document}
\draft
\title{ON THE  THREE-DIMENSIONAL TEMPORAL SPECTRUM OF STRETCHED VORTICES}
\author{Maurice Rossi}
\address{Laboratoire de Mod\'elisation en M\'ecanique, \\
Universit\'e de Paris VI, \\
4 Place Jussieu, F-75252 Paris cedex 05, France.}
\author{St\'ephane Le Diz\`es}
\address{Institut de Recherche sur les Ph\'enom\`enes Hors \'Equilibre, \\
UMR 138, CNRS et Universit\'es Aix-Marseille I \&  II, \\
12 Avenue G\'en\'eral Leclerc, F-13003 Marseille, France.}

\date{December 23, 1996}

\maketitle

\begin{abstract}
The three-dimensional stability
problem of a  stretched stationary vortex is addressed in this letter.
More specifically, we prove  that the discrete part of
the temporal spectrum is only associated with two-dimensional
perturbations.
\end{abstract}

\section{INTRODUCTION}

Numerical simulations\cite{Vincent,Jimenez,Kida} as well as  real
experiments\cite{Cadot} indicate that vorticity in turbulent flows
concentrates  in localized regions such as filaments which
are fairly well described by stretched vortices such as the celebrated Burgers
vortex  solution\cite{Saffman}
or  Moffatt, Kida \& Ohkitani's asymptotic solution\cite{Moffatt}.
If one agrees that these local structures
are important dynamical objects of the global turbulent field,
their temporal  stability  with respect
to generic perturbations should be addressed.
So far, this problem has only been studied  for Burgers' vortex
and, in that case, for purely two-dimensional
perturbations. Robinson \& Saffman\cite{Robinson} provided
an analytical solution for low-Reynolds numbers and
 their result were later numerically extended by Prochazka \&
Pullin\cite{Prochazka}
up to  Reynolds numbers $Re =10^4$.
These papers on Burgers  vortex indicate that
the temporal spectrum associated with two-dimensional
perturbations is discrete and corresponds to damped
modes.
On  the contrary, the  general
stability problem of stretched vortices has not been tackled yet.
Except for the 2D stability analysis of axisymmetric Burgers vortex,
 it does not reduce to a classical
eigenvalue problem with a single ODE to solve.
Indeed,  infinitesimal 3D  perturbations
are affected by the presence of stretching along the
vortex axis which precludes the reduction of the problem
by Fourier analysis.

In this paper, we prove that the discrete part of
the temporal spectrum is only associated with two-dimensional
perturbations.  In section \ref{sec:formu}, the stability problem is
introduced and  particular time-dependent solutions are
exhibited. Their existence  imposes conditions on
the 3D temporal mode structure.
In section \ref{sec:z}, these conditions are shown to
be consistent only for modes independent of the vortex axis coordinate.

\section{Modified Fourier decomposition}
\label{sec:formu}

Let us consider a stationary velocity field
${\mathbf U}_0 = (U_0,V_0,W_0)$ of the form
\begin{mathletters}
\begin{eqnarray}
U_0 = \frac{\partial \phi}{\partial x} (x,y)  + U_v(x,y)~,
\label{exp:U0}
\\
V_0 =  \frac{\partial \phi}{\partial y}(x,y) + V_v(x,y)~,
\\
W_0 = \gamma z~,
\end{eqnarray}
\end{mathletters}
where $(U_v, V_v)$ and   $ ( \frac{\partial \phi}{\partial x}(x,y),
\frac{\partial \phi}{\partial
y}(x,y), \gamma z) $  respectively stand for a localized rotational field
and a global velocity
field satisfying  $\nabla^2 \phi =- \gamma$. Such an  expression
represents a stationary stretched
vortex aligned with the $z$-axis and  subjected to a global strain field.
In the
sequel, the  strain rate $\gamma$  along the vortex axis  is assumed to be
positive.

The structure of such a solution is governed by the balance
between stretching due to the global strain field
and viscous diffusion. In particular,
the core size should scale as $\sqrt{\nu /\gamma}$ where $\nu$ is the kinematic
viscosity. In the simplest case of an axisymmetric strain  $ (
\frac{\partial \phi}{\partial
x}(x,y), \frac{\partial \phi}{\partial y}(x,y), \gamma z) =(-\gamma
x/2,-\gamma y/2,\gamma z)$, one recovers  the Burgers solution.   Other examples  are
Robinson \&  Saffman
\cite{Robinson} and  Moffatt, Kida \&  Ohkitani \cite{Moffatt} solutions
which correspond to the   non-axisymmetric case at small and large Reynolds
numbers respectively.
In the subsequent analysis, expression (\ref{exp:U0}-c) is considered as the
basic flow where $(U_v, V_v)$ and $\phi$ are not specified.

The dynamics of pressure and velocity infinitesimal
perturbations $({\mathbf u},p)$
around (\ref{exp:U0}-c) is described by the linear system~:
\begin{mathletters}
\begin{eqnarray}
\partial _t{\mathbf u}  +
{\mathbf U}_0.{\mathbf \nabla}{\mathbf u}
+ ({\mathbf u}.{\mathbf \nabla}){\mathbf U}_0
= -{\mathbf \nabla} p
+ \nu  \nabla ^2 {\mathbf u} ~,
\label{equ:u} \\
{\mathbf \nabla}.{\mathbf u} =0~.
\end{eqnarray}
\end{mathletters}

The above system being homogeneous with respect to time, one might look at
the temporal spectrum
of such system. Modes belonging to the discrete part of this  spectrum  read:
\begin{equation} ({\mathbf u}_{\omega},p_{\omega}) = ({\mathbf
v}_{\omega}(x,y,z),
q_{\omega}(x,y,z))  e^{-i\omega t}~.
\label{exp:uom}
\end{equation}
Inserting (\ref{exp:uom}) into (\ref{equ:u}-b), one obtains
equations for $ {\mathbf v}_{\omega}(x,y,z)$ and $ q_{\omega}(x,y,z)$
which are non-separable with respect to any spatial variable. In such a case,
the use of standard Fourier analysis does not simplify any further the problem.
However, the z-dependence in equation (\ref{equ:u}) only appears through
the uniform  strain  along
the $z$-axis i.e. through the term $\gamma z \partial _z$. Such a simple
dependence allows
to search for  time-dependent solutions which are different from
(\ref{exp:uom}).
These are ``generalized Fourier modes'' in the  $z$ direction with a
time-dependent wavenumber~:
\begin{equation}
({\mathbf u},p) = (\check{\mathbf u}(x,y,t,k_0), \check{p}(x,y,t,k_0))
e^{ik(t)z}
e^{-\nu \int_0^t (k(s))^2ds}~,
\label{exp:uP}
\end{equation}
where the initial wavenumber condition  $k_0 =k(0)$ is a free parameter.
Indeed, as soon as the time evolution of  $k(t)$
is appropriately chosen, more precisely if $k(t)= k_0e^{-\gamma t}$,
the non-homogeneous term in $z$ is removed in (\ref{equ:u}).
The system (\ref{equ:u},b) is then reduced to a couple
of equations homogeneous  in $z$, which describes the time evolution
of the modified Fourier modes components  $\check{u} _x$ and $\check{u} _y$~:
\begin{mathletters}
\begin{eqnarray}
({\cal L} + \partial_x U_0 )\check{u} _x + \partial_y U_0 \check{u} _y =
\frac{e^{2\gamma t}}{k_0^2}
\frac{\partial }{\partial x}({\cal L} + 2\gamma)\left(\frac{\partial
\check{u} _x}{\partial x} + \frac{\partial \check{u} _y}{\partial y}\right)
~,
\label{equ:cux} \\
({\cal L} + \partial_y V_0 )\check{u} _y + \partial_x V_0 \check{u} _x =
\frac{e^{2\gamma t}}{k_0^2}
\frac{\partial }{\partial x}({\cal L} + 2\gamma)\left(\frac{\partial
\check{u} _x}{\partial x} + \frac{\partial \check{u} _y}{\partial y}\right)
~,
\end{eqnarray} \end{mathletters}
where
\begin{equation}
{\cal L} = \frac{\partial }{\partial t} +  U_0 \frac{\partial }{\partial
x}+ V_0 \frac{\partial }{\partial y}
- \nu \left( \frac{\partial ^2 }{\partial x^2} +\frac{\partial ^2
}{\partial y^2} \right)~.
\label{exp:bL}
\end{equation}
Finally, note that $\check{u} _z$ is given by the components  $\check{u}
_x$ and $\check{u} _y$
through  the continuity equation.

The temporal mode (\ref{exp:uom}) may be decomposed upon a basis of such
modified Fourier modes
(\ref{exp:uP}).  Let us expand  the spatial part $({\mathbf
v}_{\omega},q_{\omega})$ of
(\ref{exp:uom})  in the usual Fourier modes along the $z$ direction~:
\begin{equation}
({\mathbf v}_{\omega},q_{\omega}) =
\int_{-\infty}^{+\infty} ({\mathbf v}(x,y,k_0), q(x,y,k_0))
e^{ik_0z}dk_0~.
\label{exp:uF}
\end{equation}
This expansion can be viewed as a superposition at time $t=0$
of generalized Fourier modes  provided the following initial
condition is satisfied~:
\begin{equation}
(\check{{\mathbf u}}(x,y,0,k_0), \check{p} (x,y,0,k_0))
= ({\mathbf v}(x,y,k_0), p(x,y,k_0))~.
\end{equation}
According to (\ref{exp:uP}), each   generalized Fourier mode evolves
independently: an
alternative expression for the temporal mode (\ref{exp:uom}) is thus
provided for all $t$
\begin{equation}
({\mathbf u}_{\omega},p_{\omega}) = \int_{-\infty}^{+\infty}
(\check{{\mathbf u}}(x,y,t,k_0), \check{p} (x,y,t,k_0)) e^{ik_0e^{-\gamma t}z}
e^{-\nu k^2_0 (1-e^{-2\gamma t})/2\gamma} dk_0~.
\end{equation}
The above expression  is   consistent with (\ref{exp:uom}) if the following
equality holds  at
any $x$ and $y$ locations, time $t$ and  wavenumber $k_0$~:
\begin{equation} (\check{{\mathbf
u}}(x,y,t,k_0), \check{p} (x,y,t,k_0)) e^{-\nu k_0^2 (1-e^{-2\gamma t})/2\gamma}
= ({\mathbf v}(x,y,k_0e^{-\gamma t}), q(x,y,k_0e^{-\gamma t}))e^{-\gamma t}
e^{-i\omega t}.
\label{exp:uP=uF}
\end{equation}
In the following section,  equality (\ref{exp:uP=uF}) is shown to be valid
only for $k_0 =
0$: discrete temporal modes are bound to be two-dimensional.

\section{The $z$ dependence of a three-dimensional temporal mode}
\label{sec:z}

First  a cut-off wavenumber $k_c$  above which ${\mathbf v}(x,y,k_0)$ and
$q (x,y,k_0)$ vanish,
should exist. Indeed, assume that such a cut-off does not appear, large
wavenumbers are then
present in the spatial spectrum  of (\ref{exp:uom}). It is thus possible to take
the simultaneous  limits $k_0 t $ large and  $t$ small  in
(\ref{exp:uP=uF}). The right-hand side of this equation becomes
\begin{equation}
({\mathbf v}(x,y,k_0e^{-\gamma t}), q(x,y,k_0e^{-\gamma t}))e^{-\gamma t}
e^{-i\omega  t}
\sim ({\mathbf v}(x,y,k_0), q(x,y,k_0)) ~.
\label{exp:LHS}
\end{equation}
In order to  estimate the left-hand side of (\ref{exp:uP=uF}), the behavior
of $\check{{\mathbf u}}(x,y,t,k_0),
\check{p} (x,y,t,k_0))$  is evaluated using  equations (\ref{equ:cux},b).
Two cases are to  be
considered according to the characteristic spatial variations  of
$\check{u} _x$ and $\check{u} _y$
in the $x$ and $y$ directions.  When, for large   axial wavenumber  $k_0$,
these components evolve over
spatial scales independent of $k_0$, the right-hand side of
(\ref{equ:cux},b) can be neglected and
the leading order time evolution is independent on $k_0$. This means that,
for   $k_0t$ large and
$t$ small,  the left-hand side of (\ref{exp:uP=uF})  reads
\begin{equation}
(\check{{\mathbf u}}(x,y,t,k_0), \check{p} (x,y,t,k_0))
e^{-\nu k_0^2 (1-e^{-2\gamma t})/2\gamma}
\sim
(\check{{\mathbf u}}_{\infty}(x,y,0), \check{p} _{\infty} (x,y,0))
e^{-\nu k_0^2 t}~.
\label{exp:RHS1}
\end{equation}
On the contrary, when  $\check{u} _x$, $\check{u} _y$ evolve over spatial scales
comparable to $1/k_0$, the correct expression is found  when introducing in
(\ref{equ:cux},b)
new  fast variables    \begin{equation}
\overline{x} = k_0 x~; ~ \overline{y} = k_0 y ~;~ \overline{t} = k_0^2 t,
\end{equation}
in addition to $x$ and $y$, and thereafter expanding the solution with
respect to $1/k_0$. At
leading order, homogeneous equations are obtained~:
\begin{mathletters}
\begin{eqnarray}
\frac{\partial \check{u} _x}{\partial \overline{t}}
- \nu \left( \frac{\partial ^2 }{\partial \overline{x}^2} +\frac{\partial
^2 }{\partial \overline{y}^2} \right)\check{u} _x =
e^{2\gamma t} \left[
\frac{\partial }{\partial \overline{t}}
- \nu \left( \frac{\partial ^2 }{\partial \overline{x}^2} +\frac{\partial
^2 }{\partial \overline{y}^2} \right)\right]
\frac{\partial }{\partial \overline{x}}\left(\frac{\partial \check{u}
_x}{\partial \overline{x}} + \frac{\partial \check{u} _y}{\partial
\overline{y}}\right) ~,
\label{equ:cuxinf} \\
\frac{\partial \check{u} _x}{\partial \overline{t}}
- \nu \left( \frac{\partial ^2 }{\partial \overline{x}^2} +\frac{\partial
^2 }{\partial \overline{y}^2} \right)\check{u} _y =
e^{2\gamma t} \left[
\frac{\partial }{\partial \overline{t}}
- \nu \left( \frac{\partial ^2 }{\partial \overline{x}^2} +\frac{\partial
^2 }{\partial \overline{y}^2} \right)\right]
\frac{\partial }{\partial \overline{x}}\left(\frac{\partial \check{u}
_x}{\partial \overline{x}} + \frac{\partial \check{u} _y}{\partial
\overline{y}}\right) ~.
\end{eqnarray} \end{mathletters}

For $t$ small, the term $e^{2\gamma t}$ may be  taken equal to 1 and  the
general solution of (\ref{equ:cuxinf})  be written in the fast variables
$\overline{x}$ and
$\overline{y}$ via the usual Fourier decomposition
 \begin{equation}
(\check{u} _x, \check{u} _y) = \int
(\check{u} _x^{(0)}(k_x,k_y,x,y),\check{u} _y^{(0)}(k_x,k_y,x,y))
 e^{-\nu ( k_x^2 + k_y^2)\overline{t} } e^{ik_x \overline{x} +ik_y
\overline{y}}dk_x dk_y~.
\label{exp:cuxcuy}
\end{equation}
For fixed $x$ and $y$ and  $k_0 t \rightarrow +\infty$,  the above
expression (\ref{exp:cuxcuy})
is   evaluated by the steepest descent method
\begin{equation}
(\check{u} _x, \check{u} _y) \sim \frac{4 \pi}{\nu k_0^2 t}
(\check{u} _x^{(0)}(0,0,x,y),\check{u} _y^{(0)}(0,0,x,y))~.
\label{exp:cuxcuyl}
\end{equation}
Similar behaviors can be  obtained for $\check{u} _z$ and $\check{p}$. For
this case, the
right-hand side  of (\ref{exp:uP=uF}) now reads
\begin{equation}
(\check{{\mathbf u}}(x,y,t,k_0), \check{p} (x,y,t,k_0))
e^{-\nu k_0^2 (1-e^{-2\gamma t})/2\gamma}
\sim
(\check{{\mathbf u}}^{(0)}(0,0, x,y), \check{p} ^{(0)}(0,0,x,y))
\frac{4\pi }{\nu k_0^2 t}e^{-\nu k_0^2 t}~.
\label{exp:RHS2}
\end{equation}

Introducing into equation (\ref{exp:uP=uF}),  estimates (\ref{exp:LHS}) and
(\ref{exp:RHS1}) or (\ref{exp:RHS2}) leads to an inconsistency for large
wavenumbers
$k_0$. This contradiction  implies that  the spatial spectrum  of
(\ref{exp:uom}) vanishes for
sufficiently  large wavenumbers~:  there hence exists  a cut-off wavenumber
$k_c$ such that
$({\mathbf v}(x,y,k_0), q(x,y,k_0)) =0$  for $k_0 > k_c$.

Consider now a wavenumber $0<k_1 < k_c$ such that   $({\mathbf v}(x,y,k_1),
q(x,y,k_1)) \neq
({\mathbf 0},0)$ and  a time $t_1$ such that $k_0= k_1e^{\gamma t_1} > k_c$.
Equation (\ref{exp:uP=uF}) then implies that  $(\check{{\mathbf
u}}(x,y,t_1,k_0), \check{p}
(x,y,t_1,k_0)) \neq 0$ which leads again to a contradiction. Indeed,   the
initial
condition   $(\check{{\mathbf u}}(x,y,0,k_0), \check{p} (x,y,0,k_0)) $ is
equal to zero  since $k_0>k_c$, the quantity  
$(\check{{\mathbf u}}(x,y,t,k_0), \check{p}(x,y,t,k_0))$  thus remains  zero
for all $t>0$ since it is   governed by equation  (\ref{equ:cux},b).  This
imposes that   $({\mathbf
v}(x,y,k_1), q(x,y,k_1)) =({\mathbf 0},0)$   for all $ k_1 \neq 0$.
Discrete temporal modes are then
independent upon $z$ i.e.   purely two-dimensional ones.   For the
axisymmetric Burgers vortex,
those have been computed  by Robinson  \&  Saffman\cite{Robinson} and
Prochazka and
Pullin\cite{Prochazka}. These authors showed that these modes  are damped.

Needless to say, the  general stability problem for Burgers vortex or any
stretched vortex,  cannot
be  solved  only looking at the discrete part of the temporal spectrum. In
particular, the analysis
of the continuous spectrum need to be considered.  However, the process of
``bidimensionalization''
displayed  in (\ref{exp:uP}) can be used again in that case.


\begin{thebibliography}{99}

\bibitem{Vincent} A. Vincent and M. Meneguzzi, ``The spatial structure and
statistical properties of homogeneous turbulence,'' J. Fluid Mech. {\bf
225}, 1 (1991).


\bibitem{Jimenez} J. Jim\'enez, A. A. Wray, P. G. Saffman and R. S.
Rogallo, J. Fluid Mech. {\bf 255}, 65 (1993).

\bibitem{Kida} S. Kida, ``Tube-like structures in turbulence,'' Lecture
Notes in Numerical Applied Analysis {\bf 12}, 137 (1993).


\bibitem{Cadot} O. Cadot, S. Douady and Y. Couder,
``Characterization of the low  pressure filaments in
three-dimensional turbulent shear flow,'' Phys. Fluids
{\bf 7}, (3), 630
(1995).

\bibitem{Saffman} P. G. Saffman, {\it Vortex Dynamics} (Cambridge
University Press, Cambridge, 1992).


\bibitem{Moffatt}  H. K. Moffatt, S. Kida and K. Ohkitani, ``Stretched
vortices$-$the sinews of turbulence; large-Reynolds-number asymptotics,''
J. Fluid Mech. {\bf 259}, 241 (1994).


\bibitem{Robinson} A. C. Robinson and P. G. Saffman, ``Stability and Structure
of Stretched Vortices,'' Stud. Appl. Math. {\bf 70}, 163
(1984).

\bibitem{Prochazka} A. Prochazka and D. I. Pullin, ``On the two-dimensional
stability of the axisymmetric Burgers vortex,'' Phys. Fluids {\bf 7}, 1788
(1995).




\end{thebibliography}
\end{document}